\documentclass[psfig]{mn2e}

\usepackage{times}

\newcommand{\bgr}{\bibitem[\protect\citename{dummy }1893]{dum}}

\begin{document}
\title[Absorbed QSO hosts]
{Submillimetre photometry of X-ray absorbed QSOs: their formation and evolutionary status}

\author[J. A. Stevens et al.]
{J.\ A.\ Stevens,$^{1,2}$ M.\ J.\ Page,$^3$ R.\ J.\ Ivison,$^{2,4}$ 
F.\ J.\ Carrera,$^{5}$ J.\ P.\ D.\ Mittaz,$^{6}$ Ian\ Smail$^7$
\newauthor and I.\ M.\ McHardy$^{8}$ 
\\
$^1$ Centre for Astrophysics Research, Science and Technology Research Centre,
University of Hertfordshire, College Lane, Herts, AL10 9AB\\
$^2$ UK Astronomy Technology Centre, Royal Observatory, Blackford Hill,
Edinburgh EH9 3HJ \\
$^3$ Mullard Space Science Laboratory, University College London, Holmbury
St. Mary, Dorking, Surrey, RH5 6NT\\
$^4$ Institute for Astronomy, University of Edinburgh, Royal Observatory,
Edinburgh EH9 3HJ \\ 
$^5$ Instituto de Fisica de Cantabria (CSIC-UC), Avenida de los Castros
39005 Santander, Spain\\
$^6$ Department of Physics, University of Alabama in Huntsville, Huntsville, AL
35899, USA\\
$^7$ Institute for Computational Cosmology, University of Durham, South Road,
Durham  DH1 3LE\\ 
$^8$ Department of Physics and Astronomy, University of Southampton,
Southampton, SO17 1BJ
}
\date{draft 1.0}

\maketitle
\begin{abstract}
\noindent 
We present an analysis of the submillimetre/X-ray properties of 19 X-ray
absorbed, Compton-thin quasi-stellar objects (QSOs) selected to have
luminosities and redshifts which represent the peak of cosmic QSO
activity. i.e. $\sim L^*$ objects at $1<z<3$.  Of these, we present new data
for 11 objects not previously observed at submillimetre wavelengths and
additional data for a further 3. The detection rate is $42$ per cent, much
higher than typically reported for samples of QSOs.  Detection statistics show
(at the $3-4\sigma$ level) that this sample of absorbed QSOs has a higher
submillimetre output than a matched sample of unabsorbed QSOs. We argue that
the far-infrared luminosity is produced by massive star formation. In this
case, the correlation found between far-infrared luminosity and redshift
can be interpreted as cosmological evolution of the star-formation rate in the
QSO host galaxies.  Since the submillimetre luminous phase is confined to
$z>1.5$, the high star formation rates are consistent with a scenario in which
the QSOs evolve to become local luminous elliptical galaxies.

Combining these results with previously published data for X-ray
unabsorbed QSOs and submillimetre-selected galaxies we propose the following
evolutionary sequence: the forming galaxy is initially far-infrared luminous
but X-ray weak similar to the sources discovered by the Submillimetre
Common-User Bolometer Array; as the black hole and spheroid grow with time a
point is reached when the central QSO becomes powerful enough to terminate the
star formation and eject the bulk of the fuel supply (the Compton-thin absorbed
QSO phase); this transition is followed by a period of unobscured QSO activity
which subsequently declines to leave a quiescent spheroidal galaxy.
\end{abstract}

\begin{keywords}
galaxies - formation: galaxies - evolution: X-rays - galaxies
\end{keywords}

\footnotetext{E-mail: jas@star.herts.ac.uk}
%

\begin{table*}
\centering
\caption[dum]{\small Measured redshifts, X-ray fluxes, absorbing column densities and
submillimetre flux densities. Also given are the coordinates of the optical QSO.}
\label{table:data}
\vspace*{0.1in}
\begin{tabular}{lccccccc}\hline
\def\baselinestretch{1.5}                                     
Source &      RA  & Dec & z & $S_X(\times10^{-14})^a$& Log($N_{\rm H}$) &
$S_{850}^b$ & $S_{450}^b$ \\
& (J2000.0) & (J2000.0) & &  (erg\,cm$^{-2}$\,s$^{-1}$)  &
(cm$^{-2}$) & (mJy) & (mJy) \\ \hline
RX\,J$005734.78-272827.4$ & $00\ 57\ 34.94$ & $-27\ 28\ 28.0$ & $2.19$ &
$1.99^{+0.75}_{-0.56}$ &
$22.65^{+0.21}_{-0.50}$ & $11.7\pm1.2$ & $51\pm9$ \\ 
RX\,J$033340.22-391833.4$ & $03\ 33\ 39.54$ & $-39\ 18\ 41.4$ & $1.44$ &
$6.68^{+2.63}_{-2.15}$ &
$22.49^{+0.16}_{-0.41}$ & $0.9\pm1.6$ & \ldots \\ 
RX\,J$092103.75+621504.3$ & $09\ 21\ 02.88$ & $+62\ 15\ 06.9$ & $1.46$ &
$5.10^{+1.51}_{-1.31}$ &
$22.11^{+0.26}_{-0.45}$ & $0.7\pm1.4$ &
$18\pm12$ \\
RX\,J$094144.51+385434.8$ & $09\ 41\ 44.61$ & $+38\ 54\ 39.1$ & $1.82$ &
$2.98^{+0.91}_{-0.83}$ &
$21.92^{+0.46}_{-0.42}$ &
$13.4\pm1.5^{c}$ & $74\pm11^{c}$ \\ 
RX\,J$094356.53+164244.1$ & $09\ 43\ 57.30$ & $+16\ 42\ 49.9$ & $1.92$ &
$8.68^{+4.37}_{-2.91}$ &
$22.72^{+0.21}_{-0.38}$ & $3.0\pm1.2$ &
$29\pm10$ \\
XMM\,J$100205.31+554258.9$ & $10\ 02\ 05.35$ & $+55\ 42\ 57.6$ & $1.15$ &
$5.43^{+1.32}_{-1.15}$ &
$22.84^{+0.18}_{-0.19}$ &
$1.5\pm0.9$ & $12\pm8$ \\
RX\,J$101033.47+533922.5$ & $10\ 10\ 33.69$ & $+53\ 39\ 29.0$ & $2.46$
&$2.82^{+1.06}_{-0.77}$&
$22.29^{+0.41}_{-0.68}$ & $-1.6\pm1.2$ &
$-1\pm9$ \\
RX\,J$101112.05+554451.3$ & $10\ 11\ 12.30$ & $+55\ 44\ 47.0$ & $1.25$ &
$8.90^{+1.94}_{-1.31}$ &
$21.62^{+0.43}_{-0.18}$ & $2.6\pm1.0$ & $6\pm6$ \\ 
RX\,J$101123.17+524912.4$ & $10\ 11\ 22.67$ & $+52\ 49\ 12.3$ & $1.01$ &
$8.67^{+4.35}_{-2.68}$ &
$22.51^{+0.15}_{-0.29}$ & $-1.0\pm1.0^{d}$ & $-5\pm9$\\ 
RX\,J$104723.37+540412.6$ & $10\ 47\ 23.50$ & $+54\ 04\ 06.7$ & $1.50$ &
$2.90^{+1.72}_{-0.99}$ &
$22.22^{+0.41}_{-0.78}$ & $1.7\pm1.1^{d}$ & $-9\pm8$\\  
RX\,J$110431.75+355208.5$ & $11\ 04\ 32.44$ & $+35\ 52\ 14.1$ & $1.63$ &
$5.43^{+2.65}_{-1.59}$ &
$22.26^{+0.32}_{-0.68}$ & $2.4\pm1.2$ &
$29\pm11$ \\
RX\,J$110742.05+723236.0$ & $11\ 07\ 41.59$ & $+72\ 32\ 35.8$ & $2.10$ &
$10.53^{+2.45}_{-1.66}$ &
$21.58^{+0.82}_{-0.37}$ &
$10.4\pm1.2$ & $8\pm9$ \\
RX\,J$111942.16+211518.1$ & $11\ 19\ 42.13$ & $+21\ 15\ 16.6$ & $1.29$ &
$4.19^{+0.82}_{-0.68}$ &
$21.42^{+0.36}_{-0.29}$ &
$-0.5\pm1.0^{d}$ & $13\pm8$ \\
RX\,J$121803.82+470854.6$ & $12\ 18\ 04.54$ & $+47\ 08\ 51.0$ & $1.74$ &
$2.36^{+1.15}_{-0.68}$ &
$22.30^{+0.35}_{-0.56}$ &
$6.8\pm1.2^{e}$ & $24\pm9$\\  
RX\,J$124913.86-055906.2$ & $12\ 49\ 13.85$ & $-05\ 59\ 19.4$ & $2.21$ &
$3.39^{+0.85}_{-0.84}$ &
$22.23^{+0.35}_{-0.63}$ & 
$7.2\pm1.4^{e}$ & $81\pm12$ \\
XMM\,J$133447.34+375950.9$ & $13\ 34\ 47.40$ & $+37\ 59\ 50.0$ & $1.18$ &
$1.48^{+0.07}_{-0.07}$ &
$21.83^{+0.06}_{-0.07}$ & 
$0.9\pm1.3$ & $-4\pm13$ \\
RX\,J$135529.59+182413.6$ & $13\ 55\ 29.54$ & $+18\ 24\ 21.3$ & $1.20$ &
$6.25^{+2.84}_{-1.72}$ &
$22.24^{+0.24}_{-0.50}$ &
$0.8\pm1.3^{e}$ & $-8\pm10$\\
RX\,J$140416.61+541618.2$ & $14\ 04\ 16.79$ & $+54\ 16\ 14.6$ & $1.41$ &
$3.97^{+0.77}_{-0.99}$ &
$21.69^{+0.39}_{-0.39}$ &
$0.1\pm1.1$ & $11\pm12$ \\
RX\,J$163308.57+570258.7$ & $16\ 33\ 08.59$ & $+57\ 02\ 54.8$ & $2.80$ &
$2.66^{+0.74}_{-0.58}$ &
$22.47^{+0.32}_{-0.48}$ &
$5.9\pm1.1^{e}$ & $9\pm9$ \\ \hline
\multicolumn{8}{l}{\small $^a$ Absorption corrected $0.5-2$~keV X-ray flux}\\
\multicolumn{8}{l}{\small $^b$ Quoted errors do not include calibration
uncertainties (see Section~\ref{section:red})}\\
\multicolumn{8}{l}{\small $^c$ Flux densities from Stevens et al. (2004).
The quoted 450~$\mu$m value corresponds to emission from source 1 and 2 in
that paper.} \\
\multicolumn{8}{l}{\small $^d$ New data combined with those from Page et al. (2001)}\\
\multicolumn{8}{l}{\small $^e$ From Page et al. (2001)}
\end{tabular}
\end{table*}

\section{Introduction}
In the last decade research into active galactic nuclei has moved from the
periphery of astrophysics to take centre stage in the study of galaxy
formation. This role change was triggered by the discovery that most local
spheroidal galaxy components (elliptical galaxies and the bulges of spiral
galaxies) contain a massive black hole with mass proportional to that of the
velocity dispersion of the stars (Magorrian et al. 1998) or equivalently the
mass of the spheroid (e.g. McLure \& Dunlop 2002). These relationships suggest
a direct link between the growth of the black hole and the stellar mass of the
galaxy spheroid.

A second major advance in our knowledge of galaxy formation has occurred in
parallel with that involving nuclear activity: the discovery made with the
Submillimetre Common-user Bolometer Array (SCUBA; Holland et al. 1999) on the
James Clerk Maxwell Telescope (JCMT) of a large population of ultraluminous
star-forming galaxies at $z>1$ (e.g. Smail, Ivison \& Blain 1997; Hughes et
al. 1998). Semi-analytic models have struggled to reproduce the number counts
of these massive galaxies at high redshifts (e.g. Devriendt \& Guiderdoni 2000;
Baugh et al. 2004) although more recent schemes which include feedback from the
active nucleus appear to be more promising (Granato et al. 2004).

There are two obvious direct approaches that can be taken to investigate the
coupled formation of a galactic bulge and its black hole. One is to make X-ray
observations of known submillimetre-bright galaxies. The first experiments
showed little overlap between the two populations (Fabian et
al. 2000). However, once more sensitive X-ray observations became available
many SCUBA-selected galaxies were found to have weak X-ray counterparts
(Alexander et al. 2003, 2004). From these data, it can be concluded that the
major episode of star formation does not coincide with the epoch of visible
quasi-stellar object (QSO) activity although an evolutionary link between
SCUBA-selected galaxies and QSOs may still exist (e.g. Almaini 2003).

The second approach is that taken here: to select typical AGN at X-ray
wavelengths, i.e. those responsible for the majority of the comoving accretion
luminosity density, and observe them at submillimetre wavelengths (Page et
al. 2001). To date we have observed two matched samples of X-ray selected
QSOs. The first of these comprised 8 QSOs showing strong evidence for
photoelectric absorption in their X-ray spectra -- indicating the presence of
gas in the line-of-sight to the nucleus. Of these, the 4 at $z>1.5$ were
detected as ultraluminous or hyperluminous far-infrared galaxies while the 4 at
$z<1.5$ were not. We argued that these results were consistent with the coeval
growth of the galaxy spheroid and black hole in these systems (Page et
al. 2001) but that, with the available data, it was not possible to determine
whether the redshift trend could be attributed to cosmological evolution or
selection effects. The second sample comprised 20 objects that showed no
evidence for photoelectric absorption; these have properties akin to the
majority of QSOs selected at optical wavelengths. Only one of these QSOs was
detected by SCUBA with $>3\sigma$ significance (Page et al. 2004).

This striking result indicates that the absorption in these objects is not a
consequence of the basic `unified scheme' (Antonucci 1993) in which the X-ray
properties of QSOs are determined by viewing angle alone. Instead it can be
argued that the absorbed QSOs are transition objects caught between an epoch of
hidden growth and an unobscured QSO phase, and that the absorbed phase
coincides with the formation of the galaxy spheroid (Page et al. 2004).

The motivation for the present study is straightforward: X-ray absorbed QSOs
selected by our criteria must be important objects in the history of galaxy
formation but only 8 objects have been observed at submillimetre
wavelengths. Here we report sensitive submillimetre observations of another 11
objects making the sample sufficiently large to (1) improve the detection
statistics, and (2) search for trends between submillimetre and X-ray
properties.  A Hubble constant $H_0=70$\ km\,s$^{-1}$\,Mpc$^{-1}$ and density
parameters $\Omega_{\Lambda}=0.7$ and $\Omega_{\rm m}=0.3$ are assumed
throughout.

\begin{table*}
\centering
\caption[dum]{\small X-ray and far-infrared luminosities calculated from the
data presented in Table~\ref{table:data}. We also give the ratio of
far-infrared luminosity to bolometric luminosity of the QSO (see text for
details) and to $0.5-2.0$ keV X-ray luminosity.}
\label{table:lums}
\vspace*{0.1in}
\begin{tabular}{lcccc}\hline
Source & Log($L_X)$ & Log($L_{\rm FIR}$) & ($L_{\rm FIR}/L_{\rm QSO})$ & ($L_{\rm FIR}/L_{\rm X})$\\
& (erg\ s$^{-1}$) & (L$_{\odot}$) & & \\ \hline
RX\,J$005734.78-272827.4$ & $44.85^{+0.14}_{-0.14}$ & $13.32^{+0.08}_{-0.06}$ &
$3.4\pm1.4$ & $114\pm47$ \\ 
RX\,J$033340.22-391833.4$ & $44.93^{+0.15}_{-0.17}$ & $<12.90$ & $<1.1$ & $<36$ \\ 
RX\,J$092103.75+621504.3$ & $44.83^{+0.11}_{-0.13}$ & $<12.84$ & $<1.2$ & $<39$\\
RX\,J$094144.51+385434.8$ & $44.83^{+0.12}_{-0.14}$ & $13.41^{+0.05}_{-0.07}$ &
$4.8\pm1.8$ & $146\pm55$ \\ 
RX\,J$094356.53+164244.1$ & $45.36^{+0.17}_{-0.18}$ & $12.76^{+0.13}_{-0.22}$ &
$0.3\pm0.2$ & $10\pm7$ \\
XMM\,J$100205.31+554258.9$ & $44.61^{+0.09}_{-0.11}$ & $<12.79$ & $<1.7$ & $<58$ \\
RX\,J$101033.47+533922.5$ & $45.13^{+0.14}_{-0.14}$ & $<12.80$ & $<0.5$ & $<18$ \\
RX\,J$101112.05+554451.3$ & $44.90^{+0.09}_{-0.06}$ & $<12.92$ & $<1.2$ & $<40$ \\ 
RX\,J$101123.17+524912.4$ & $44.67^{+0.18}_{-0.16}$ & $<12.53$ & $<0.8$ & $<28$ \\ 
RX\,J$104723.37+540412.6$ & $44.62^{+0.20}_{-0.18}$ & $<12.86$ & $<2.0$ & $<69$ \\  
RX\,J$110431.75+355208.5$ & $44.98^{+0.17}_{-0.15}$ & $12.65^{+0.18}_{-0.30}$ &
$0.5\pm0.4$ & $18\pm16$ \\
RX\,J$110742.05+723236.0$ & $45.54^{+0.09}_{-0.08}$ & $13.28^{+0.06}_{-0.08}$ &
$0.6\pm0.2$ & $21\pm6$ \\
RX\,J$111942.16+211518.1$ & $44.61^{+0.08}_{-0.07}$ & $<12.56$ & $<1.0$ & $<34$ \\
RX\,J$121803.82+470854.6$ & $44.69^{+0.17}_{-0.15}$ & $13.10^{+0.08}_{-0.10}$ &
$3.0\pm1.5$ & $99\pm50$ \\  
RXJ\,J$124913.86-055906.2$ & $45.10^{+0.10}_{-0.13}$ & $13.11^{+0.09}_{-0.11}$ &
$1.2\pm0.5$ & $39\pm7$ \\
XMM\,J$133447.34+375950.9$ & $43.97^{+0.02}_{-0.03}$ & $<12.80$ & $<7.8$ & $<260$\\
RX\,J$135529.59+182413.6$ & $44.71^{+0.16}_{-0.14}$ & $<12.79$ & $<1.4$ & $<46$ \\
RX\,J$140416.61+541618.2$ & $44.68^{+0.08}_{-0.12}$ & $<12.63$ & $<1.0$ & $<34$ \\
RX\,J$163308.57+570258.7$ & $45.24^{+0.11}_{-0.10}$ & $13.00^{+0.08}_{-0.11}$ &
$0.7\pm0.2$ & $22\pm3$  \\ \hline
\end{tabular}
\end{table*}

\section{Sample selection, observations and data reduction}
\label{section:red}

We selected X-ray absorbed QSOs from the {\em Rosat\/} sample of Page, Mittaz
\& Carrera (2001) and from the new generation of surveys being conducted with
{\em XMM-Newton\/} and {\em Chandra\/} (e.g. Page et al. 2003; McHardy et
al. 2003). They are selected to (1) span the redshift range $1<z<3$ and (2)
have $0.5-2$~keV X-ray luminosities $\sim1-10$ $L_{\rm X}^*$ where $L_{\rm
X}^*$ is the break in the X-ray luminosity function measured in erg~s$^{-1}$
and has values $44.1<{\rm Log}(L_{\rm X}^*)<44.4$ for $1<z<3$ (e.g. Page et al
2004).  They are thus typical objects selected at the peak of QSO activity. The
QSOs are all Compton thin with absorbing column densities measured in cm$^{-2}$
of $21<{\rm Log}(N_{\rm H})<23$ and have broad optical emission lines.

For the {\em Rosat}-selected QSOs, column densities were determined by fitting
an absorbed power law to a three-band {\em Rosat\/} PSPC spectrum. The source and
background counts, which were determined from images constructed in the three
energy bands, are given in table~2 of Page, Mittaz \& Carrera (2000).  The
three energy bands correspond to PSPC PI channels $11-41$, $52-90$, and
$91-201$. The mean X-ray power-law spectral index of unabsorbed QSOs is
$\alpha=1$ (where $f_{\nu}\propto \nu^{-\alpha}$), with a standard deviation of
$\sim 0.2$ (Page et al. 2003; Mateos et~al. 2005). Therefore we assumed a
power-law index $\alpha=1$ and included a fixed component of Galactic
absorption in the fit.  The power-law normalization and the column density of a
cold absorber at the redshift of the QSO were free parameters in the fit, which
was performed using the Cash statistic $C$ (Cash 1979). Uncertainties on the
fit parameters were obtained first by constructing two-dimensional $\Delta C$
contours and then marginalised by integrating over one of the parameters, as
described in Mittaz et al. (1999). This is identical to the method used by
Page, Mittaz \& Carrera (2001).

\begin{figure}
\setlength{\unitlength}{1in}
\begin{picture}(3.0,5.0)
\includegraphics{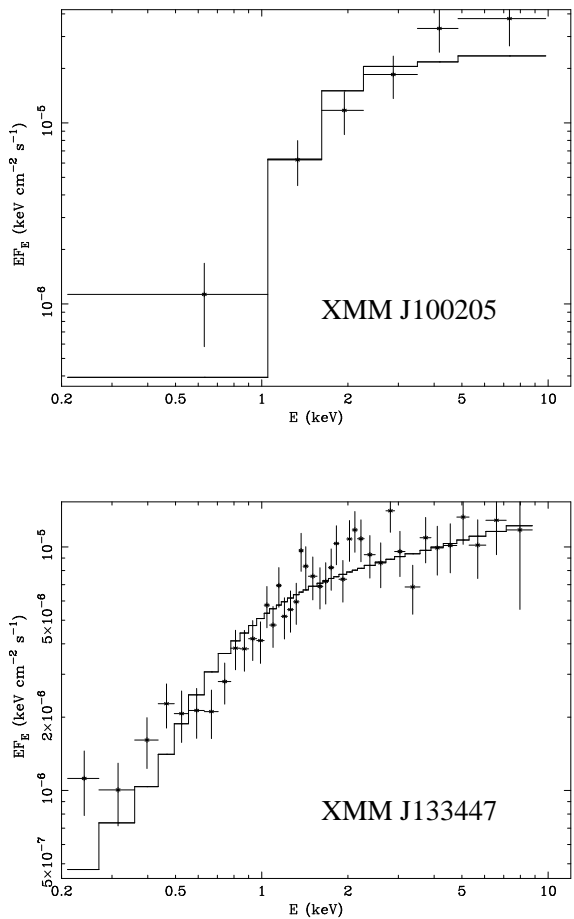}
\end{picture}
\caption[dum]{The {\em XMM-Newton\/} spectra of XMM\,J$100205+554258.9$ (top) and
XMM\,J$133447+375950.9$ (bottom) along with the best-fitting absorbed power-law
models (stepped lines).  The {\em XMM-Newton\/} effective area has been divided
out of both model and data in these plots.}
\label{fig:xmmspecs}
\end{figure}

For the two {\em XMM-Newton}-selected QSOs, X-ray spectra were constructed from
EPIC {\em pn\/} (Str\"uder et~al. 2001) and MOS (Turner et~al. 2001) data. The
data were reduced using the {\em XMM-Newton\/} Science Analysis System ({\sc
sas}) V6.0.  Source counts were obtained from elliptical spatial regions,
oriented to match the off-axis {\em XMM-Newton\/} point spread
function. Background counts were obtained from an annular region surrounding
the source, with bright sources excised. Response matrices and effective area
files were constructed using the appropriate {\sc sas} tasks. For each source,
the MOS and {\em pn\/} spectra were then combined and suitable background and
response files were constructed using the method of Page, Davis \& Salvi
(2003). Finally, the spectra of XMM\,J$100205+554258.9$ and
XMM\,J$133447+375950.9$ were grouped to minima of 16 and 30 counts per bin
respectively before $\chi^{2}$ fitting. Both objects are viewed through
extremely low Galactic absorbing columns of $8\times
10^{19}$~cm$^{-2}$. However, XMM\,J$100205+554258.9$ lies behind an additional
line of sight column density of $4.8\times 10^{20}$~cm$^{-2}$ which is
associated with NGC~3079 (Womble, Junkkarinen \& Burbidge 1992), and so for
this object we set the Galactic column density to $5.6\times 10^{20}$~cm$^{-2}$
in the spectral fit.  The X-ray spectrum of XMM\,J$100205+554258.9$ is
acceptibly fitted ($\chi^{2}/\nu=6.5/4$) with the same model as used for the
{\em Rosat\/} QSOs (i.e. an absorbed power law with fixed $\alpha=1$).
XMM\,J$133447+375950.9$, however, is not well fitted by this model. Instead,
allowing a slightly harder continuum slope of $\alpha=0.7$ (still within the
range commonly observed in unabsorbed AGN) results in an acceptable
$\chi^{2}/\nu=45/39$. The two {\em XMM-Newton\/} spectra are shown in
Fig. \ref{fig:xmmspecs}.  Names, optical coordinates, redshifts, X-ray fluxes
and absorbing column densities are listed in Table~\ref{table:data}.

Submillimetre photometry observations at 450 and 850~$\mu$m were performed at
the JCMT during 2003 December and 2004 January.  We used SCUBA in standard
2-bolometer chopping mode which allows a second bolometer on each array to
sample the source position in one half of the nod cycle giving a $(4/3)^{1/2}$
reduction in the noise compared to 1-bolometer mode. Data were taken
simultaneously at 450 and 850~$\mu$m in $\sim1$~hour blocks separated by
pointing checks on a near-by blazar and focus checks as appropriate. The sky
opacity was monitored on a quasi-continuous basis with the JCMT water vapour
radiometer and cross-checked with skydips about twice per night. Flux density
calibration was made against Uranus or the compact secondary calibrator
CRL\,618 (Sandell 1994; Jenness et al. 2002). Calibration uncertainties are
about $10$ and $15-20$ per cent at 850 and 450~$\mu$m respectively.  We aimed
for a uniform noise level of $\sim1$~mJy at 850~$\mu$m thus allowing a proper
statistical analysis of the submillimetre properties. For this reason we
re-observed 3 sources previously reported by Page et al. (2001).  Observations
were made in good-to-excellent conditions with the 225 GHz opacity as measured
at the adjacent Caltech Submillimeter Observatory always $<0.08$. During the
best weather we made 6 detections at 450~$\mu$m with signal-to-noise (S/N)
greater than or approximately 3.

Data were reduced with the {\sc starlink} {\sc surf} package in the standard
manner (see e.g. Holland et al. 1999). We reduced data from different nights
and with different bolometers separately and calculated a weighted mean flux
density and error. The final photometry results are listed in
Table~\ref{table:data}.

\section{Results}

\subsection{Luminosities}

\label{section:lums}

In Table~\ref{table:lums} we present X-ray and far-infrared luminosities for
each source. The 850-$\mu$m flux densities are converted into far-infrared
luminosites ($L_{\rm FIR}$) by scaling them with that of Mrk 231 which is a
local ultraluminous infrared galaxy (ULIRG) hosting an X-ray absorbed QSO; we
calculate $L_{\rm FIR}=1.9\times10^{12}~{\rm L}_{\rm \odot}$ by modelling its
millimetre--far-infrared spectral energy distribution. Quoted uncertainties on
$L_{\rm FIR}$ are calculated by adding in quadrature the error given in
Table~\ref{table:data} with the estimated 10 per cent calibration
uncertainty. Note that the uncertainties will be under-estimates
unless the QSOs have the same spectrum as Mrk~231. For example,
if we used Arp~220 or Mrk~273 as templates then calculated values of $L_{\rm
FIR}$ would be systematically lower by factors of 1.5 and 1.6
respectively. However, we stress that the choice of template only affects the
normalization of the derived luminosities; its variation as a function of
redshift over the range covered by our sources is small for all realistic
templates because the strong negative K-correction at 850~$\mu$m cancels out
the effect of cosmological dimming (e.g. Blain \& Longair 1993).  The template
producing the smallest variation in flux density between $1.0<z<2.8$ is Mrk~231
($\sim1$ per cent) whilst the largest variation ($\sim20$ per cent) comes from
Arp~220. The maximum scatter introduced by this effect is thus roughly
comparable to the estimated uncertainties ($\sim0.1$~dex) on the computed
luminosities and will not affect our conclusions. Upper limits to $L_{\rm FIR}$
are calculated from the 850-$\mu$m flux densities ($S_{850}$) using values
($2\sigma+S_{850}$) if $S_{850}>0$ and $2\sigma$ if $S_{850}<0$.  Bolometric
luminosities for the QSOs are calculated assuming that 3 per cent of the
bolometric output of the AGN is emitted in the $0.5-2.0$~keV band (Elvis et
al. 1994). Note that these bolometric luminosities are computed from the rest
frame spectral energy distributions of the QSOs and include data ranging from
radio to hard X-ray wavelengths. They thus include contributions from the
direct AGN continuum and reprocessed emission that emerges at longer
wavelengths.

\subsection{Detection statistics and comparison with other samples}

\label{section:det}

\begin{table*}
\centering
\caption[dum]{\small Correlation probabilities from the survival analysis
tests. Quoted values give the probability that a correlation is present between
the two parameters.}
\label{table:corr}
\vspace*{0.1in}
\begin{tabular}{llcccc}\hline
 Sample & Test & $P(L_{\rm X},z)$ & $P(L_{\rm FIR},z)$ & $P(L_{\rm FIR},L_{\rm X})$ &
$P(L_{\rm FIR},N_{\rm H})$\\ \hline
Full ($n=19$)  & Cox hazard & $0.997$ & $0.998$ & $0.979$ & $0.653$\\
               & Kendall's tau & $>0.999$ & $0.995$ & $0.917$ & $0.233$\\
Restricted ($n=13$) & Cox hazard & $0.727$ & $>0.999$ & $0.813$ & $0.558$\\
               & Kendall's tau & $0.889$ & $0.995$ & $0.635$ & $0.488$\\\hline
\end{tabular}
\end{table*}
 
With this sample of 19 sources we can place better statistical constraints on
the prevalence of luminous star formation in the absorbed QSO population than
was possible in our previous study which contained 8 sources (Page et
al. 2001).  Throughout this paper two sources with $2<{\rm (S/N)}<3$ in both
wavebands are treated as detections: RX\,J$094356.53+164244.1$ and
RX\,J$110431.75+355208.5$. Both of these objects have higher S/N at $450~\mu$m
than at $850~\mu$m which is quite common for thermal sources observed in
excellent weather conditions. We consider these sources to be `submillimetre
detections' and use their 850-$\mu$m flux densities in subsequent analysis.  A
third QSO, RX\,J$101112.05+554451.3$ is marginally `detected' at 850~$\mu$m
($2.6\sigma$) but not at 450~$\mu$m. We do not consider this source as a
detection but note that its inclusion as such would not alter the
interpretation of the correlation results presented in
Section~\ref{section:corr}. The other 6 detections are all at $>5\sigma$
significance at 850~$\mu$m. As found for the smaller sample, the detection rate
is high (42 per cent) at 850~$\mu$m. The weighted mean flux density of the
complete sample is $3.1\pm0.3$~mJy. The detected and non-detected subsamples
have weighted means of $7.3\pm0.4$ and $0.6\pm0.3$~mJy respectively.

Two of the QSOs are radio loud: RX\,J$110431.75+355208.5$ and
RX\,J$163308.57+570258.7$. Both of these objects were detected at submillimetre
wavelengths so we need to consider whether the 850~$\mu$m emission could be due
to synchrotron radiation rather than reprocessed thermal emission. The case of
RX\,J$163308.57+570258.7$ was discussed by Page et al. (2001) who extrapolated
its radio spectrum assuming a power-law and found an insignificant synchrotron
contribution of $<0.1$~mJy at 850~$\mu$m. For RX\,J$110431.75+355208.5$ the
Very Large Array FIRST survey gives an integrated 1.4-GHz flux density of
71.5~mJy while the Westerbork Northern Sky Survey gives an integrated 325-MHz
flux density of 298~mJy. A naive power-law extrapolation of these measurements
gives $\sim0.3$~mJy at $850~\mu$m.  A synchrotron contribution at $850~\mu$m
cannot thus be ruled out, although the thermal emission is likely to be a much
more significant component. Given the uncertainties we do not apply a
correction and assume that the submillimetre radiation is thermal emission from
warm dust.

This sample contains a significantly higher proportion of detections than a
similarly selected sample of unabsorbed QSOs in which we detected only 1 object
in 20 (5 per cent) with $>3\sigma$ significance (Page et al. 2004). Note that
each QSO in this unabsorbed sample was observed down to the same $\sim1$~mJy
flux density limit as the absorbed sample and none of them would qualify as a
`submillimetre detection' based on their combined 850- and 450-$\mu$m
photometry. The resulting weighted mean 850-$\mu$m flux density of these 20
unabsorbed QSOs was found to be only $0.7\pm0.2$~mJy.  Similarly, in a $z\sim2$
optically selected sample of very high luminosity ($>20L^*$) QSOs only 9 out of
57 objects (16 per cent) were detected with $>3\sigma$ significance (Priddey et
al. 2003). Although these latter submillimetre observations were relatively
shallow (the median rms flux density was $\sim2.8$~mJy), the weighted mean flux
density of the undetected sources was only $1.9\pm0.4$ mJy, insufficiently high
to rule out AGN heating as the power source of the submillimetre dust emission
(Priddey et al. 2003). These values are consistent with those found from
stacking the submillimetre flux density at the positions of faint X-ray sources
detected by {\em Chandra\/} and {\em XMM-Newton}. For the former, Barger et
al. (2001) found a noise-weighted mean 850-$\mu$m flux density of
$1.2\pm0.3$~mJy while Almaini et al. (2001) found $0.9\pm0.3$~mJy. For the
latter, Waskett et al. (2003) report $\sim 0.4 \pm 0.3$ mJy.

By contrast, in the present sample we detected 8 out of 19 QSOs. 
The Bayesian posterior probability of getting $n$ sources showing a given
property out of a sample of $N$ sources is $$P(f)\propto (f+f_{spu})^n
(1-f-f_{spu})^{(N-n)}$$ where $f$ is the fraction of the underlying population
showing that property, and $f_{spu}$ the fraction of spurious detections (in
the case of $3\sigma$ detections $f_{spu}=0.0027\equiv 1-0.9973$). This $P(f)$
can be normalized integrating over $f$ between 0 and 1; we have assumed here a
diffuse prior, meaning that the a priori probability of $f$ is flat between 0
and 1. As intuitively expected, $P(f)$ peaks at $\sim n/N-f_{spu}$, and
confidence intervals can be obtained integrating it around that peak to the
desired confidence level. For the X-ray absorbed QSOs presented here ($n=8$ and
$N=19$) between 19 and 61 per cent of the population from which they are
selected will be detected at 850~$\mu$m with a flux density $>2-3$~mJy at 95
per cent confidence. For the matched sample of unabsorbed QSO presented in Page
et al. (2004) ($n=1$ and $N=20$) less than $\sim 21$ per cent of the objects
will present that level of emission.

This formalism can also be used to compare two samples. Under the null
hypothesis that both samples are drawn from the same underlying population, in
which a fraction $f$ of sources are detectable in the submillimetre band,
$$P(f)\propto (f+f_{spu})^{(n+m)} (1-f-f_{spu})^{(N+M-n-m)}$$ where again $n$
sources are detected out of $N$ in one sample, and $m$ out of $M$ in a second
independent sample. Given this $P(f)$, if we draw from the total population two
independent samples of sizes $N$ and $M$, respectively, the probability
$P(i,j;N,M)$ of detecting $i$ sources in the first sample, and $j$ in the
second is given by
\begin{eqnarray*}
P(i,j;N,M) & = & \left({N \over i}\right)\left({M \over j}\right) \int_0^1
df\,P(f)\,(f+f_{spu})^{(i+j)}\\
 &\times &(1-f-f_{spu})^{(N+M-i-j)}.
\end{eqnarray*}
Finally, two samples are significantly different if the probability of
detecting $\geq n$ sources in the first sample and $\leq m$ sources in the
second ($P(\geq n,\leq m;N,M)=\sum_{i=n}^{N}\sum_{j=0}^{m} P(i,j;N,M)$) is
smaller than a chosen significance. We find $P(\geq 8,\leq 1;19,20)=0.0014$,
which means that our matched samples of X-ray absorbed and unabsorbed AGN are
different at $>3\sigma$ significance. If we consider the $z>1.5$ samples only,
we find $P(\geq 8,\leq 1;10,12)=5.6\times10^{-5}$ ($>4\sigma$ significance).


\begin{figure*}
\setlength{\unitlength}{1in}
\begin{picture}(5.0,3.3)
\includegraphics{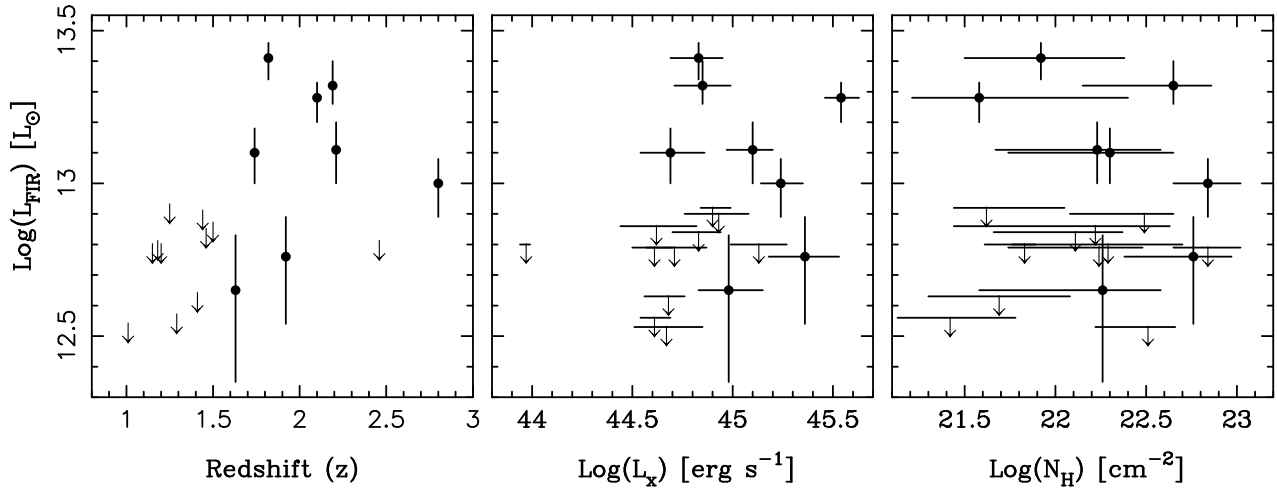}
\end{picture}
\caption[dum]{The panels (from left to right) show far-infrared luminosity
($L_{\rm FIR}$) versus redshift, $0.5-2.0$~keV X-ray luminosity ($L_{\rm X}$) and X-ray
absorbing column density ($N_{\rm H}$). 
}
\label{fig:photom}
\end{figure*}

\begin{figure*}
\setlength{\unitlength}{1in}
\begin{picture}(5.0,2.6)
\includegraphics{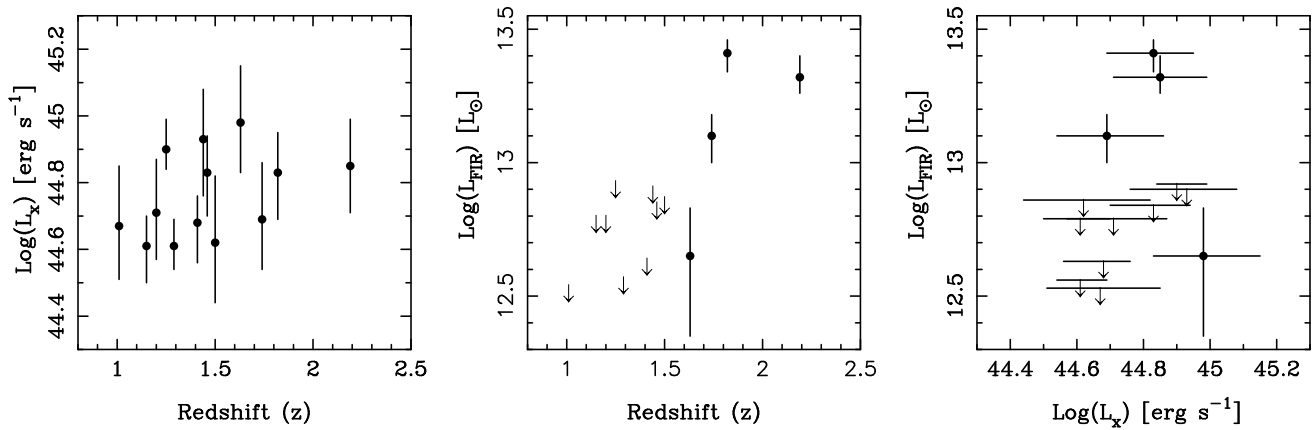}
\end{picture}
\caption[dum]{The panels (from left to right) show X-ray luminosity ($L_{\rm
X}$) versus redshift, far-infrared luminosity ($L_{\rm FIR}$) vs redshift, and
X-ray luminosity vs far-infrared luminosity for the sample of QSOs restricted
to have $44.5 \leq \rm{Log}(L_{\rm X}) \leq 45.0$.  The only significant
correlation is between $L_{\rm FIR}$ and redshift (see text for details).}
\label{fig:rest}
\end{figure*}

\subsection{Correlation analysis}

\label{section:corr}

We would like to know whether $L_{\rm FIR}$ is correlated with redshift ($z$),
X-ray luminosity ($L_{\rm X}$) or absorbing column density ($N_{\rm H}$). These
quantities are plotted against each other in Fig.~\ref{fig:photom}. Since our
dataset contains a large fraction of upper limits we have to use `survival
analysis' techniques (e.g. Isobe, Feigelson \& Nelson 1986).  The most
appropriate tests are those performed by the Cox proportional hazard model and
the Generalized Kendall's tau. Spearman's rank order test which is known to be
unreliable for datasets of $<30$ points is not used. The two adopted tests are
suitable for datasets in which the dependent variable (in our case $L_{\rm
FIR}$) contains only one kind of censoring (in our case upper limits).

Using the {\sc iraf} tasks {\sc coxhazard} and {\sc bhkmethod} we tested for
correlations between $L_{\rm FIR}$ and $z$, $L_{\rm X}$ and $N_{\rm H}$ in the
full sample of 19 QSOs. The correlation probabilities are given in
Table~\ref{table:corr}. Note that although the adopted correlation tests return
the probability $Q$ that a correlation is not present, the quoted values
$P=1-Q$ give the probability of the opposite outcome. Both tests return high
probabilities for a correlation between $L_{\rm FIR}$ and $z$ and low
probabilities for a correlation between $L_{\rm FIR}$ and $N_{\rm H}$. However,
they both return a reasonably high likelihood that $L_{\rm FIR}$ is correlated
with $L_{\rm X}$. In this case we must test whether $L_{\rm X}$ correlates with
$z$ since a trend between $L_{\rm FIR}$ and $L_{\rm X}$ could result from a
mutual correlation with $z$. Table~\ref{table:corr} shows that both tests
return a high probability that $L_{\rm X}$ and $z$ are correlated.

For the full sample of 19 QSOs this analysis shows that we cannot determine
whether $L_{\rm FIR}$ correlates with $L_{\rm X}$ or with $z$. However, when
selecting the QSOs we endeavoured to minimize any correlation between $L_{\rm
X}$ and $z$ which allows us to select a sizable subsample for which these two
parameters are uncorrelated. Thus we can restrict the sample to comprise those
13 objects satisfying $44.5 \leq {\rm Log}(L_{\rm X}) \leq 45.0$. The
correlation analysis for this restricted sample is also given in
Table~\ref{table:corr} and we plot $L_{\rm X}$ vs $z$, $L_{\rm FIR}$ vs $z$ and
$L_{\rm FIR}$ vs $L_{\rm X}$ in Fig.~\ref{fig:rest}. For these 13 QSOs there is
now no correlation between $L_{\rm X}$ and $z$ or between $L_{\rm X}$ 
and $L_{\rm FIR}$, but both tests give a strong
correlation between $L_{\rm FIR}$ and $z$. We note that the resticted range in
$L_{\rm X}$ makes the analysis insensitive to any real
correlation between $L_{\rm FIR}$ and $L_{\rm X}$; this is not the right sample
to search for such a correlation. However, the analysis does show
that the correlation between $L_{\rm FIR}$ and $z$ is real, i.e. it does not
result from a mutual correlation with $L_{\rm X}$.  We discuss the implications
of this correlation in Section~\ref{section:int}.

\subsection{The power source of the far-infrared emission}

It has already been argued that the intrinsic difference between the
submillimetre properties of absorbed and unabsorbed QSOs suggests that the
former are undergoing major starbursts (see Page et al. 2004 and
Section~1). For completeness we should also consider the viability of the
alternative scenario in which the AGN heats the dust. In this case, the
observed dichotomy in far-infrared luminosity must be related to an intrinsic
difference in the circumnuclear environment of the two types of QSOs. We note,
however, that such an interpretation appears somewhat ad hoc since it is not
obviously apparent what the physical explanation for such a difference would be
(although see Barger et al. 2005).  Nevertheless, such a possibility is
considered below.

We first consider the diagnostic used by Page et al. (2001), namely the ratio
of far-infrared luminosity to bolometric output of the QSO. These values are
listed for each QSO in Table~\ref{table:lums}.  In three cases the measured
far-infrared luminosities are significantly greater than the entire bolometric
output of the QSO, and in most other cases comprise a sizable fraction of
it. For the former cases the QSO cannot possibly power the far-infrared
luminosity.  For the latter, the QSO can only power the far-infrared emission
if the majority of the QSO emission is absorbed and reprocessed by dust. Since
we observe these QSOs to be relatively unattenuated in the rest-frame
ultraviolet, it also appears unlikely that they are the dominant power sources
for the far-infrared dust emission.

It is also instructive to compare the far-infrared/X-ray output of our QSOs
with that of the $0<z<1$ UVSX quasars presented by Elvis et al. (1994). These
are `normal' optically bright quasars that were detected in the X-ray band by
{\em Einstein}. We first removed from the sample objects classified as
flat-spectrum compact, steep spectrum compact or Fanaroff-Riley class 2 steep
spectrum doubles. Since the output from these objects is likely contaminated by
non-thermal emission they are not good lower-redshift analogues of our QSOs.
Far-infrared luminosities were calculated for the remaining 34 quasars by
scaling their 60-$\mu$m flux densities with that of IZw~1 (PG~0050+124), one of
the UVSX quasars with flux density measurements at far-infrared through
submillimetre wavelengths (Hughes et al. 1993; Haas et al. 2000). We fitted its
spectrum with an isothermal greybody giving a far-infrared luminosity of
$3.3\times10^{11}~{\rm L}_{\odot}$.  The $0.1-1.0$~keV X-ray luminosities from
Elvis et al. (1994) were converted to our adopted cosmology and corrected to
$0.5-2.0$~keV luminosities assuming a photon index, $\Gamma=2.0$.
 
\begin{figure}
\setlength{\unitlength}{1in}
\begin{picture}(3.5,2.7)
\includegraphics{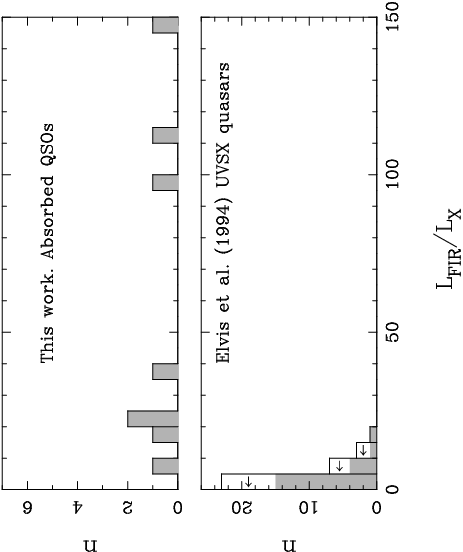}
\end{picture}
\caption[dum]{Histograms of the ratio of far-infrared luminosity to X-ray
luminosity for (top panel) SCUBA detected absorbed QSOs from this work and
(bottom panel) UVSX quasars from Elvis et al. (1994). On the bottom panel
shaded regions show the number of detections in each bin while unshaded regions
with arrows show lower limits.}
\label{fig:hist}
\end{figure}

We plot the histogram of $L_{\rm FIR}/L_{\rm X}$ in Fig.~\ref{fig:hist} (lower
panel) and show for comparison (upper panel) the histogram of values for our
SCUBA-detected absorbed QSOs (Table~\ref{table:lums}). While the number of
absorbed QSOs is small, the two histograms are clearly different. The UVSX
quasars have values that peak strongly (30/34 cases) in the first two bins
($L_{\rm FIR}/L_{\rm X}<10$) while the absorbed QSOs have values ranging
between $\sim10$ and $150$. If we assume that the X-rays are produced close to
the central engine by the same mechanism for all quasars then this result
suggests that the far-infrared emission does not have the same origin in the
two samples. For the UVSX quasars, the $L_{\rm FIR}/L_{\rm X}$ values have a
small scatter -- the 60-$\mu$m detected sources have a mean value of $5\pm5$ --
suggesting that the far-infrared luminosity is physically related to the output
of the AGN. Indeed, the survival analysis tasks {\sc bhkmethod} and {\sc
coxhazard} return correlation probabilities of $>0.999$ and $0.999$
respectively. An obvious interpretation is that the far-infrared emission is
reprocessed quasar continuum, possibly from dust in a circumnuclear torus. For
the absorbed QSOs the excess far-infrared luminosity can be attributed to dust
heated by hot young stars.

In addition, we note that recent submillimetre imaging results for one of the
absorbed QSOs show dust emission extended over tens of kpc in an apparent
merger morphology (Stevens et al. 2004). Similar observations of high-redshift
radio galaxies often yield the same result (Stevens et al. 2003). For the
submillimetre-detected absorbed QSOs we conclude that all available evidence
points towards dust heated in a major starburst rather than by an AGN
continuum.

Finally, let us consider the submillimetre non-detected absorbed QSOs.  In
Section~\ref{section:det} we found that the mean flux density of these objects
is $0.6\pm0.3$ mJy. Scaling with IZw~1 which has a predicted 850-$\mu$m flux
density of $0.34-0.39$~mJy over the relevant redshift range ($1.0<z<1.5$) gives
a mean far-infrared luminosity of $\sim6\pm3\times10^{11}~{\rm L}_{\odot}$. The
mean X-ray luminosity for these objects using the values from
Table~\ref{table:lums} is $5.0^{+4.4}_{-2.4}\times10^{44}~{\rm erg\
s}^{-1}$. Thus the mean value of $L_{\rm FIR}/L_{\rm X}\sim4\pm4$ for the
non-detected absorbed QSOs is wholly consistent with the mean value found for
the UVSX quasars suggesting that they may have similar properties.

\section{Interpretation: An evolutionary sequence?}

\label{section:int}

In the previous section it was argued that the submillimetre emission from
SCUBA-detected absorbed QSOs can be attributed to dust heated by hot young
stars. In this case, the implied star-formation rates, given by ${\rm SFR}({\rm
M_{\odot}}\,{\rm yr}^{-1})=L_{\rm FIR}/(5.8\times10^9{\rm L}_{\rm \odot}$)
(Kennicutt 1998) are $>1000~{\rm M_{\odot}}\,{\rm yr}^{-1}$, sufficiently high
to build a substantial fraction of a galaxy spheroid in only a few 100 Myr.  In
Page et al. (2004) we argued (1) that their space density relative to
unabsorbed QSOs indicates that the absorbed phase has a duration $\sim15$ per
cent that of the unabsorbed phase (2) that unabsorbed QSOs have already built
most of their stellar mass implying that the absorbed phase precedes the epoch
of luminous unobscured QSO activity (3) that the X-ray luminosities of the
absorbed sources imply that they contain black holes of mass $>10^8~{\rm
M_{\odot}}$. The conclusion of that work was thus that absorbed QSOs are
observed during a transition phase between their highly-obscured growth and a
period of unabsorbed QSO activity.

The analysis of a larger sample of absorbed QSOs presented here strengthens
these claims. In particular, we have shown statistically that absorbed QSOs
have higher submillimetre flux densities than unabsorbed QSOs of given X-ray
luminosity and redshift. Their submillimetre detection rate is also much higher
than luminous QSOs selected at optical wavelengths.  In addition, for the
restricted sample, we find evidence that X-ray absorbed QSOs were forming stars
much more rapidly at early epochs. All 8 QSOs detected at submillimetre
wavelengths lie at $z>1.5$. This result indicates cosmological evolution of the
star-formation rate density in the absorbed QSO population. A similar result
was found for radio galaxies (Archibald et al. 2001; Reuland et al. 2004)
although these have a much lower space density than $\sim L_{\rm X}^*$
QSOs. Such an early epoch of copious star formation is consistent with these
QSOs evolving to become elliptical galaxies at lower redshifts. Indeed, at
$0<z<0.5$ we know that elliptical galaxies are preferentially found in cluster
environments where they have old, coeval stellar populations. The small
intrinsic scatter in the colours of ellipticals within these clusters suggests
that they formed the bulk of their stars synchronously at $z>2$ or so (Ellis et
al. 1997). This result is explained naturally by the hierarchical model of
galaxy formation since density peaks on galaxy scales collapse, on average, at
higher redshifts if they are situated in large-scale over densities. Such
over-dense regions evolve to form a cluster at $z=0$ (Kauffmann 1996).


We can compare our proposed evolutionary sequence for QSOs with the physical
model presented by Granato et al. (2004) who consider joint spheroid/QSO
evolution within a hierarchical clustering scenario (although see Baugh et
al. 2004 who discuss some of the assumed simplifications). Their model, which
is able to reproduce the SCUBA number counts, makes a number of predictions
that are in good accord with the results presented here and by Page et
al. (2004). It predicts that the luminous starburst phase begins before the
black hole has built sufficient mass to shine as a luminous QSO, but that this
starburst is still ongoing, with a ${\rm SFR}>1000~{\rm M}_{\rm \odot}{\rm
yr}^{-1}$, as the black hole reaches its final mass of $10^8-10^9~{\rm M}_{\rm
\odot}$ (for a $10^{12.4}-10^{13.4}~{\rm M}_{\rm \odot}$ dark matter
halo). Sources caught in this phase would have properties commensurate with
those of our $z>1.5$ X-ray absorbed QSOs.  During this transition the quasar
feedback removes most of the gas and dust leaving the nucleus shining as an
unabsorbed/optically-selected QSO (see also Silk \& Rees 1998; Fabian 1999; Di
Matteo, Springel \& Hernquist 2005) -- consistent with the lack of evidence for
substantial star formation in such objects discussed earlier. Sources caught
during the initial luminous starburst phase may comprise the bulk of the
submillimetre galaxy population discovered with SCUBA (Smail, Ivison \& Blain
1997) -- many are ULIRGs containing growing black holes with Seyfert-like X-ray
luminosities (Alexander et al. 2003; Smail et al. 2003; Alexander et al. 2004)
and they often show signatures of buried AGN in their optical spectra (Chapman
et al. 2003, 2004).



Where do the $z<1.5$ absorbed QSOs fit into this scheme?  None of these objects
are luminous submillimetre sources. One possibility is that the absorption in
the $z<1.5$ objects arises in an obscuring circumnuclear torus whereas at
$z>1.5$ it is related to the major formation epoch of the stellar
spheroid. Such a possibility, while plausible, is a little contrived. A second,
and maybe more natural explanation, might be that fuelling of the
AGN/starbursts evolves with redshift (as already alluded to above). At high
redshifts, merger events or interactions that trigger the activity occur
between gas-rich galaxies (e.g. Kauffmann \& Haehnelt 2000), producing luminous
starbursts and obscured X-ray sources. However, at lower redshifts, while the
QSO may undergo recurrent merger induced accretion episodes that drive enough
gas into the nuclear region to produce the observed X-ray absorption, there is
insufficient fuel for them to be accompanied by starbursts luminous enough to
be detected by SCUBA.

The proposed evolutionary scenario can be investigated further by (1) measuring
the evolutionary status of the stellar spheroid and (2) characterising the
Mpc-scale environments. How do their molecular gas masses and stellar masses
compare with those of SCUBA-selected galaxies (Frayer et al. 1998, 1999; Neri
et al. 2003; Greve et al. 2005)? Are they located in over-densities of other
luminous star-forming galaxies hosting buried AGN as expected if they are to
evolve into cluster elliptical galaxies (e.g. Stevens et al. 2004)?  To answer
these questions we are now conducting observations of these absorbed QSO fields
at submillimetre, mid-infrared and X-ray wavelengths.



\section*{ACKNOWLEDGMENTS}

The James Clerk Maxwell Telescope is operated by the Joint Astronomy Centre in
Hilo, Hawaii on behalf of the parent organizations PPARC in the United Kingdom,
the National Research Council of Canada and The Netherlands Organization for
Scientific Research.  Based on observations obtained with {\em XMM-Newton\/},
an ESA science mission with instruments and contributions directly funded by
ESA member states and the USA (NASA). We thank the JCMT telescope system
specialists for their help, and in particular Jim Hoge for his expertise and
energy.  J.A.S. thanks PPARC's Rolling Grant Panel for their sagacity.
F.J.C. acknowledges support from the Spanish Ministerio de Cienca y
Technolog\'{\i}a, under project ESP2003-00812.  I.R.S. acknowledges support
from the Royal Society.

\bsp

\end{document}